\documentclass[aps,pra,twocolumn]{revtex4-1}
\usepackage{amsmath}
\usepackage{amsfonts}
\usepackage{amssymb}
\usepackage{graphicx}
\usepackage[colorlinks,linkcolor=blue,anchorcolor=blue,citecolor=blue,urlcolor=blue]{hyperref}
\usepackage{mathrsfs}
\usepackage{epsfig}
\usepackage{float}
\usepackage{subfigure}
\usepackage{mathtools} 
\begin{document}

\title{Localization transitions of correlated particles in nonreciprocal quasicrystals}

\author{Lei Wang$^{1}$} 
\author{Juan Kang$^{2,3}$} 
\author{Ni Liu$^{4}$} 
\thanks{Corresponding author: liuni2011520@sxu.edu.cn}
\author{Chaohua Wu$^{2,3}$}
\thanks{Corresponding author: sxwuchua@163.com}
\author{Gang Chen$^{2,3}$} 
\thanks{Corresponding author: chengang971@163.com}

\affiliation{$^{1}$State Key Laboratory of Quantum Optics and Quantum Optics Devices, Institute of Laser spectroscopy, Shanxi University, Taiyuan 030006, China}
\affiliation{$^{2}$Laboratory of Zhongyuan Light, School of Physics, Zhengzhou University, 
Zhengzhou 450001, China}
\affiliation{$^{3}$Key Laboratory of Materials Physics, Ministry of Education, School of Physics,
Zhengzhou University, Zhengzhou 450001, China}
\affiliation{$^{4}$Institute of Theoretical Physics, Shanxi University, Taiyuan 030006, China}

\date{\today}

\begin{abstract}

The interplay among interaction, non-Hermiticity, and disorder opens a new avenue for engineering novel phase transitions. We here study the spectral and localization features of two interacting bosons in one-dimensional nonreciprocal quasicrystals. Specifically, by considering a quasiperiodic Hubbard lattice with nonreciprocal hoppings, we show that the interaction can lead to a mobility edge, which arises from the fact that the bound states display a much lower threshold for spectral and extended-localized transitions than scattering states. The localization transition of bound or scattering states is accompanied by a complex-real spectrum transition. Moreover, while the two-particle localized states are robust to the boundary conditions, the two-particle extended states turn into skin modes under open boundary condition. We also show the correlated dynamics to characterize these localization transitions. Finally, we reveal that the bound states can form mobility edge on their own by introducing a dimerized nonreciprocal quasicrystal. Our paper may pave the way for the study of non-Hermitian few-body physics.

\end{abstract}

\maketitle

\section{Introduction}

 Non-Hermitian systems with gain and loss or nonreciprocity have garnered great interest recently in many disciplines, ranging from open quantum systems to classical systems \cite{RMP015005,NM18,NP20}. Fundamentally different from Hermitian cases, non-Hermitian Hamiltonians host complex eigenvalues with nonorthogonal eigenstates and introduce novel concepts such as exceptional points \cite{Scienceeaar7709}, spectral topology \cite{PRX031079,Nature59,Science1240}, and non-Hermitian skin effects \cite{PRL086803,PRL126402,NC5491,PRL066404,PRL226402,PRL086801,PRL186802,PRL070401,PRA106}. In this context, an emerging interplay of non-Hermiticity and localization has exhibited an interesting synergy among multiple phase transitions. A paradigmatic example is the non-Hermitian quasicrystal provided by the Aubry-Andr\'{e}-Harper (AAH) Hamiltonian, showing the concurrence of the delocalization-localization transition, the parity-time symmetry breaking, and the topological transition of the eigenenergy at a critical value of the quasiperiodic potential strength \cite{PRL237601,PRB054301,PRB125157,PRB014201,PRA043325,PRA033325,PRB024203,Nature354,PRB134203}. Subsequently, there have been considerable efforts in exploring non-Hermitian localization of various generalized AAH models, such as non-Hermitian mobility edge \cite{PRB024205,PRR033052,PRB174205,PRL113601,PRB134208} and reentrant localization \cite{CPB097202,PRL106803,NJP123048,PRB1054204,PRB054307}. These studies are mostly restricted to single particle cases.

Along with the above progress, the non-Hermitian physics has been extended to interacting many-body cases. Many-body interactions trigger novel manifestations beyond single-particle picture, which brings many intriguing phenomena in non-Hermitian systems, such as many-body skin effect \cite{PRL180401,PRB235151,PRBL121102,PRBL220205,PRR033122,PRR033173,PRL076502,PRL136503,PRL096501,PRB081115} and entanglement phase transition \cite{PRX021007,PRB024306,PRA042208,PRB214308}. Particularly, the non-Hermitian many-body localization has also been investigated in disorder and disorder-free systems  \cite{PRL090603,PRB184205,PRA043301,PRB064206}, revealing that non-Hermitian many-body localization transition might coincide with or differ from the real-complex transition point.
Among strongly correlated systems, the simplest but nevertheless rich scenario to study the role of interactions is the few-body regime. One of the simplest few-body interacting models is the Bose-Hubbard model, where the few-body state can be conveniently prepared and probed in various experimental platforms such as cold atoms in optical lattices \cite{Nature853}, arrays of superconducting qubits \cite{PRB224520,Science753,PRA2409}, and classical emulators of Fock space with photonic lattices \cite{NC1555,PRA053853}. Typically, two particles on a lattice can form a bound state (doublon) for strong interactions, resulting in exotic doublon dynamics \cite{PRL085302,NP316,PRL267602,Science1229,NP1471,NAT546}.

Advances in experimental techniques, such as ultracold atoms \cite{NP385,PRL0401} and non-Hermitian topolectrical circuits \cite{NC1436,PRB195131,NC98}, have made it possible to realize non-Hermitian few-body Hamiltonian in the experiment. In this context, investigating few-body non-Hermitian quasicrystals not only deepens the understanding of peculiar localization phenomena induced by non-Hermiticity and interactions, but also offers valuable insights into the exploration of many-body non-Hermitian effects. Recently, the two-particle localization in non-Hermitian quasicrystals with complex on-site potentials has been investigated \cite{PRB075121,PRB054204}. It has been shown that the interaction-induced bound states have a lower threshold for real-complex spectral and localization transitions than single-particle states. As nonreciprocal hopping usually leads to non-Hermitian skin effects, a natural question is what localization and spectral transitions can be induced by strongly-correlated particles in nonreciprocal quasicrystals. Moreover, could the skin states and localized states can coexist and how they are related to the two-particle eigenstates?

To address the above issues, in this work we study the localization transitions of two interacting particles on a AAH lattice with asymmetric hoppings, i.e., nonreciprocal quasicrystals. By identifying the spectral and localization features of the eigenstates, we demonstrate that a mobility edge can be induced by interaction. This mobility edge arises from the fact that the interaction have different dressing effects on different eigenstates, resulting in an intermediate phase with the coexistence of extended scattering states and localized bound states. Furthermore, we show that the two-particle extended states turn into skin modes under open boundary condition, whereas the two-particle localized states are robust to different boundary conditions. The correlated dynamics are also demonstrated to characterize the localization transitions. Finally, by considering a nonreciprocal quasicrystal with dimerized hoppings, we reveal that the bound states can form mobility edge on their own. Our results provide a firm ground for understanding the many-body effects in non-Hermitian systems.

\section{Nonreciprocal interacting AAH model}
\label{SII}

\subsection{Model}

We consider a one-dimensional Hubbard model for bosons with an external
incommensurate potential modulation and asymmetric hoppings, which is viewed
as a nonreciprocal interacting AAH model. The Hamiltonian describing this
system reads \cite{PRB054301}
\begin{eqnarray}
\hat{H} &=&\sum\nolimits_{j}^{L}\left[ J(e^{\alpha }\hat{a}_{j+1}^{\dagger }%
\hat{a}_{j}+e^{-\alpha }\hat{a}_{j}^{\dagger }\hat{a}_{j+1})\right]   \notag
\\
&&+\sum_{j}^{L}2V\cos \left( 2\pi \beta j\right) \hat{n}_{j}+\frac{U}{2}%
\sum_{j}^{L}\hat{n}_{j}\left( \hat{n}_{j}-1\right) ,  \label{Eq1}
\end{eqnarray}%
where $\hat{a}_{j}$ ($\hat{a}_{j}^{\dagger }$) is the bosonic annihilation
(creation) operator at the lattice site $j$, $\hat{n}_{j}=\hat{a}%
_{j}^{\dagger }\hat{a}_{j}$ is the corresponding site number operator, and $L
$ is the length of the lattice. The first term in the above Hamiltonian
describes single-particle nonreciprocal hoppings between nearest-neighbor
sites, where the nonreciprocal strength is controlled by $\alpha $. The
second term of the Hamiltonian is the quasiperiodic modulation with
amplitude $V$. The modulation period is incommensurate with the lattice
space and characterized by an irrational number $\beta $. Without loss of
generality, we take $\beta =( \sqrt{5}-1) /2$ in our
calculations, which can be approached by $\beta =\lim\limits_{n\rightarrow
\infty }F_{n}/F_{n+1}$ with $F_{n}$ being the Fibonacci numbers defined by $%
F_{0}=F_{1}=1$ and $F_{n+1}=F_{n}+F_{n-1}$. The last term of the Hamiltonian
(\ref{Eq1}) describes the on-site interaction with strength $U$ between two bosons.
Here, we are going to focus on the case of two interacting bosons for
repulsive interaction ($U>0$). Similar analysis and analogous results can
be performed for the attractive interaction ($U<0$).

\begin{figure}[t]
\centering
\includegraphics[width=8cm]{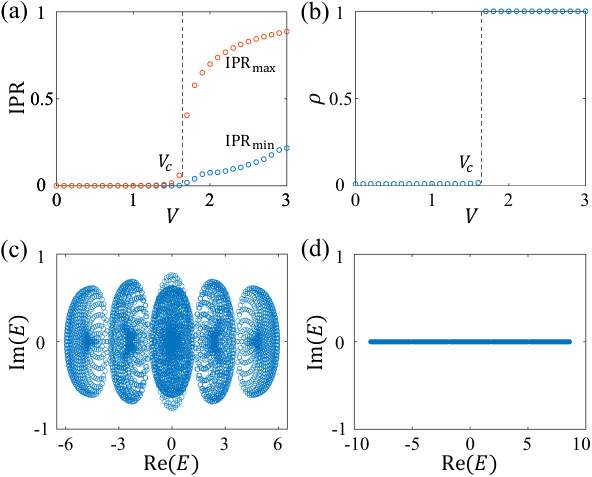}\newline
\caption{(a) The maximum (red circles) and minimum (blue circles) values of IPR of two-particle eigenstates as
functions of $V$. The dashed line corresponds to the localization transition
point $V_{c}=Je^{\left\vert \alpha \right\vert }\approx 1.65$. (b) Plot of $%
\rho $ as a function of $V$. (c, d) Two-particle complex energy spectra of
the non-Hermitian Hamiltonian~(\ref{Eq1}) with $V=1.2$ (c) and $V=2$ (d),
respectively. The other parameters are given by $\alpha =0.5$, $U=0$, and $%
L=89$.}
\label{fig1}
\end{figure}

\begin{figure*}[t]
\centering
\includegraphics[width=1.9\columnwidth]{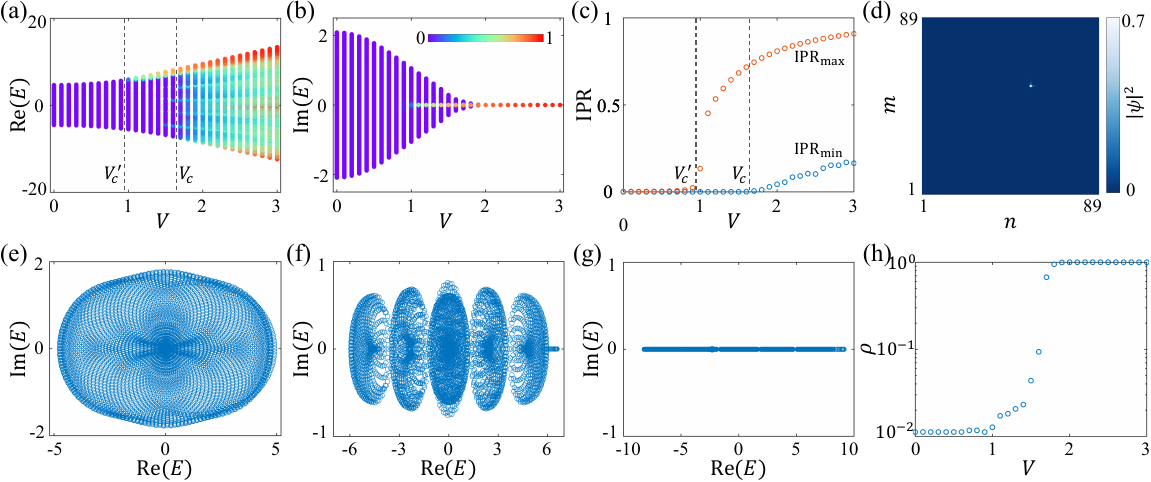}
\caption{The real (a) and imaginary (b) parts of the two-particle energy spectra of
the Hamiltonian~(\ref{Eq1}) as functions of $V$. The color code indicates the values
of IPR$_{\left( l\right) }$. The dashed lines correspond to the transition
points $V_{c}^{\prime}$ and $V_{c}$. (c) The maximum and minimum values of IPR of
two-particle eigenstates as functions of $V$. (d) Two-particle density
distribution of one localized states for $V=1.2$. (e-g) Two-particle energy
spectra of the non-Hermitian Hamiltonian (1) with $V=0.5$ (e), $V=1.2$ (f),
and $V=2$ (g), respectively. (h) Plot of $\rho $ as a function of $V$. The
other parameters are given by $\alpha =0.5$, $U=1$, and $L=89$.}
\label{fig2}
\end{figure*}

For the single particle case and $V=0$, the Hamiltonian~(\ref{Eq1}) describes the
well-known Hatano-Nelson model \cite{PRL77}, which is a paradigmatic model featuring the
non-Hermitian skin effect and non-trivial point-gap topology. In the case of
two particles, the two-body wave function can be presented in the form%
\begin{equation}
\left\vert \psi \right\rangle =\sum_{m,n}^{L}(1+\delta
_{m,n})^{-1/2}\psi _{mn}\hat{a}_{m}^{\dagger }\hat{a}_{n}^{\dagger
}\left\vert 0\right\rangle ,  \label{Eq2}
\end{equation}
where $\left\vert 0\right\rangle $ is the vacuum state, $\delta _{m,n}=1$ ($=0$) for $%
m=n$ ($m\neq n$), and $\psi _{mn}$ is the
superposition coefficients representing the probability amplitude for one
particle to be present at site $m$ and the other located at site $n$, with $%
\psi _{mn}=\psi _{nm}$ for bosons. According to Hamiltonian~(\ref{Eq1}) and Eq.~(\ref{Eq2}),
we can obtain a linear system of equations with respect to the unknown
coefficients $\psi _{mn}$, i.e.,%
\begin{eqnarray}
E \psi _{mn} &=&Je^{\alpha }\left( \psi _{m+1,n}+\psi
_{m,n+1}\right)   \notag \\
&&+Je^{-\alpha }\left( \psi _{m-1,n}+\psi _{m,n-1}\right)   \notag \\
&&+2V\left[ \cos \left( 2\pi \beta m\right) +\cos \left( 2\pi \beta n\right) %
\right] \psi _{mn}  \notag \\
&&+U\delta _{m,n}\psi _{mn},  \label{Eq3}
\end{eqnarray}
where $E$ denotes the eigenenergies. Equation~(\ref{Eq3}) indicates that the two-boson problem in a
one-dimensional nonreciprocal quasicrystal is equivalent to a
single-particle problem in a two-dimensional square lattice with an
incommensurate on-site potential and nonreciprocal hoppings. Especially, the
interaction $U$ corresponds to the on-site potential for the diagonal sites (%
$m,m$). 

In the following numerical simulation, we set $J=1$ as the energy (time) unit without losing generality.
In addition, the periodic boundary condition (PBC) is assumed unless otherwise specified.

\subsection{Noninteracting case}
Before showing the role of the interaction, we first concentrate on the
localization behaviors of two particles in the noninteracting limit $U=0$.
The two-particle eigenstates and eigenenergies can be obtained by diagonalizing
the $L(L+1)/2$ matrix Hamiltonian corresponding to $\hat{H}$ in the Fock space of two bosons. Let $\left\vert \psi _{l}\right\rangle $ be the $l$-th right
eigenstate with eigenenergy $E_{l}$, we consider the inverse participation
ratio (IPR) to describe quantitatively its localization properties,
which is defined as%
\begin{equation}
\text{IPR}_{\left( l\right) }=\sum\nolimits_{m,n=1}^{L}\left\vert \psi
_{m,n}^{\left( l\right) }\right\vert ^{4}
\label{Eq4}
\end{equation}%
with $m\geqslant n$. Note that the IPR for the left and right eigenstates behave in a similar fashion. In the limit of $L\rightarrow \infty $, the IPR of an extended state tends
to be zero, while it is finite for a localized state with IPR $<1$. In
addition, we can also define normalized participation ratio of an eigenstate
as NPR$_{\left( l\right) }=$IPR$_{\left( l\right) }^{-1}/D$ with $D$ being
the Fock-space dimension of two bosons. In contrast to the IPR, the NPR
tends to be nonzero (zero) for the extended states (localized states). To
characterize the localization features of each possible phase in our
two-particle system, we further introduce the maximum and minimum values of
IPR over the eigenstates of the Hamiltonian, which are respectively given by \textcolor{blue}{\cite{PRB054204}}
\begin{equation}
\text{IPR}_{\max }=\max\limits_{l\in \left\{ 1,\cdots ,D\right\} }\text{IPR}_{\left(
l\right) },
\end{equation}
\begin{equation}
\text{IPR}_{\min }=\min\limits_{l\in \left\{ 1,\cdots ,D\right\} }
\text{IPR}_{\left( l\right) }. 
\end{equation}
Figure~\ref{fig1}(a) shows the IPR$_{\max }$ and IPR$_{\min
}$ as functions of the quasiperiodic potential amplitude $V$ for $U=0$. One
can find a rather abrupt increase of both IPR$_{\max }$ and IPR$_{\min }
$ from zero, indicating that an extended-localized transition occurs.
Similar to the single-particle limit of Hamiltonian (\ref{Eq1}), the critical value
of this transition is given by \cite{PRB054301}%
\begin{equation}
V_{c}=Je^{\left\vert \alpha \right\vert },
\label{Eq5}
\end{equation}%
which corresponds to the large hopping of the system.

As demonstrated in the single-particle nonreciprocal AAH model \cite{PRB054301,PRA033325}, the
delocalized-localized transition is usually accompanied by a complex-real
eigenenergy transition. This feature can be confirmed by investigating the
ratio $\rho =D_{0}/D$ with $D_{0}$ being the number of eigenenergies with
zero imaginary parts. In Fig.~\ref{fig1}(b), we plot $\rho $ as a function of $V$
for two noninteracting bosons Hamiltonian (\ref{Eq1}). Clearly, for $V<V_{c}$,
almost all states have complex eigenenergies corresponding to $\rho
\approx 0$, whereas for $V>V_{c}$ the eigenenergies of all eigenstates
become real, thus yielding $\rho =1$. Figures~\ref{fig1}(c) and~\ref{fig1}(d) show the
typical energy spectra in the complex plane for $V=1.2$ and $V=2$,
respectively. It can be seen that all eigenenergies are complex (purely
real) in the extended phase (localized phase). The energy spectrum in the extended phase displays a structure of five clusters, each of which is filled by complex eigenenergies. Note that the
 number of clusters depends on $V$, which is detailed in Appendix \ref{Ap}.

\subsection{Interaction-induced spectral and localization transitions}

We now turn to the two interacting bosons case with $U\neq 0$. In Figs.~\ref{fig2}(a) and \ref{fig2}(b), we
respectively plot the real and imaginary parts of the two-particle
eigenenergies, together with their corresponding IPR$_{\left( l\right) }$,
as functions of $V$. Inspecting the values of IPRs, three distinct phases
can be clearly identified --- extended phase ($V<V_{c}^{\prime }$) with
vanishing IPRs, localized phase ($V>V_{c}$) with finite IPRs, and an
intermediate phase ($V_{c}^{\prime }<V<V_{c}$) with both vanishing and
finite IPRs. From this definition, the intermediate phase is nothing but a
parameter region where the extended and localized states coexist, leading to
the emergence of mobility edge.
To make this point clearer, we plot the IPR$%
_{\max }$ and IPR$_{\min }$ as functions of $V$, as shown in Fig.~\ref{fig2}(c).
Apart from the extended and localized phases, an intermediate region takes
place where IPR$_{\max }$ takes finite values and IPR$_{\min }$ remains to
zero (see Appendix \ref{AFS} for the finite size analysis). Remarkably, the localized states in the intermediate phase correspond
to the two-body bound states (doublons), i.e., pairs of bound particles
occupying the same site \cite{AP535}. Figure~\ref{fig2}(d) shows the two-particle density
distribution of one localized state for $V=1.2$. It is found that the
two-particle eigenstates are occupied in an individual lattice site, implying
the formation of the doublon state.

Figures~\ref{fig2}(e)-\ref{fig2}(g) show three typical energy spectra in the $V<V_{c}^{\prime }$, 
$V_{c}^{\prime }<V<V_{c}$, and $V>V_{c}$ phases, respectively. As one can notice, while all the eigenenergies are purely real in the localized
phase, only part of the eigenenergies corresponding to two-particle bound
states [Figs. \ref{fig2}(a) and \ref{fig2}(b)] are real in the intermediate phase. As shown in
Fig. \ref{fig2}(h), $\rho$ is depicted as a function of $V$. It is found that the
critical points separating different values of $\rho$ perfectly reproduce
the boundaries between phases with different localization properties.

\begin{figure}[t]
\centering
\includegraphics[width=8cm]{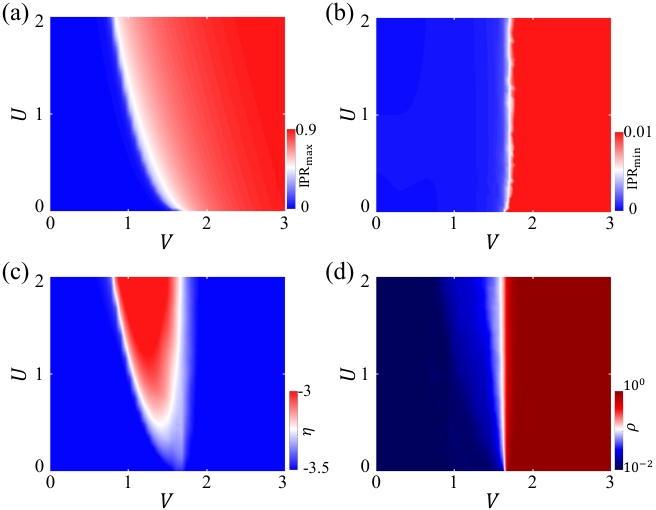}\newline
\caption{Phase diagrams of IPR$_{\max }$ (a), IPR$_{\min }$ (b), $\eta $ (c), and $%
\rho $ (d) in the $V$-$U$ plane. The other parameters are the same as those
in Fig.~\ref{fig2}.}
\label{fig3}
\end{figure}

In Figs.~\ref{fig3}(a) and \ref{fig3}(b), we plot IPR$_{\max }$ and IPR$_{\min }$ in the $V$-$U$
plane, respectively. Clearly, the boundaries of the extended-intermediate
and intermediate-localized transitions can be respectively determined by IPR$%
_{\max }$ and IPR$_{\min }$. Note that the phase boundary to localized
region (IPR$_{\min }$) is independent of the interaction $U$. The
intermediate phase can also be conveniently captured by the quantity \cite{NJP123048}%
\begin{equation}
\eta =\log _{10}\left( \text{IPR}_{\max }\times \text{IPR}_{\min }\right) .
\label{Eq6}
\end{equation}%
As displayed in Fig.~\ref{fig3}(c), we present phase diagram of $\eta $ in the $V$-$U$
plane. The presence of the intermediate region (red region) is clearly
distinguished from the fully extended or the fully localized regions (blue
regions). In Fig.~\ref{fig3}(d), we also demonstrate the characterization of
different localization phases using $\rho $. The obtained phase boundary is
consistent with the results shown in Fig. \ref{fig3}(c).

As discussed above, the intermediate phase with mobility edge shows the
coexistence of localized bound states and extended single-particle states.
To understand the physics behind this phenomenon, attention can be paid to
the strong-interaction limit $U\gg J,V$. In this case, the high-energy sector
of the energy spectrum describes doublon states with energy of order $U$,
which is well separated from the two-particle scattering states. By treating
the hopping term and on-site potential as perturbation, the effective doublon Hamiltonian to the
second-order correction can be obtained as (see Appendix \ref{Ap_A} for details)%
\begin{eqnarray}
\hat{H}_{d} &=&\frac{2J^{2}}{U}\sum\nolimits_{j}^{L}\left[ \left( e^{2\alpha
}\hat{d}_{j+1}^{\dagger }\hat{d}_{j}+e^{-2\alpha }\hat{d}_{j}^{\dagger }\hat{%
d}_{j+1}\right) \right]  \notag \\
&&+\sum\nolimits_{j}^{L}4V\cos \left( 2\pi \beta j\right) \hat{n}_{j}^{d}, 
\label{Eq7}
\end{eqnarray}%
where $\hat{d}_{j}=(1/\sqrt{2})\hat{a}_{j}^{2}$ ($\hat{d}_{j}^{\dagger }$)\ is the
doublon annihilation (creation) operator at site $j$ and $\hat{n}_{j}^{d}=%
\hat{d}_{j}^{\dagger }\hat{d}_{j}$ is the doublon number operator. From
Hamiltonian (\ref{Eq7}), the effective nonreciprocal hopping amplitude $
(2J^{2}/U) e^{\left\vert 2\alpha \right\vert }$ for the doublon is
reduced compared to the hopping amplitude $Je^{\left\vert \alpha \right\vert
}$ for single-particle case in the strong-interaction case, which is
equivalent to the increase of the amplitude of the quasiperiodic modulation.
As a result, the doublon states first undergo localization with the
increase of $V$. Concretely, the delocalized-localized transition point of
the doublon states is given by%
\begin{equation}
V_{c}^{\prime }=\frac{J^{2}}{U}e^{\left\vert 2\alpha \right\vert }=\frac{%
V_{c}^{2}}{U}.
\label{Eq8}
\end{equation}%

\begin{figure*}[t]
\centering
\includegraphics[width=1.9\columnwidth]{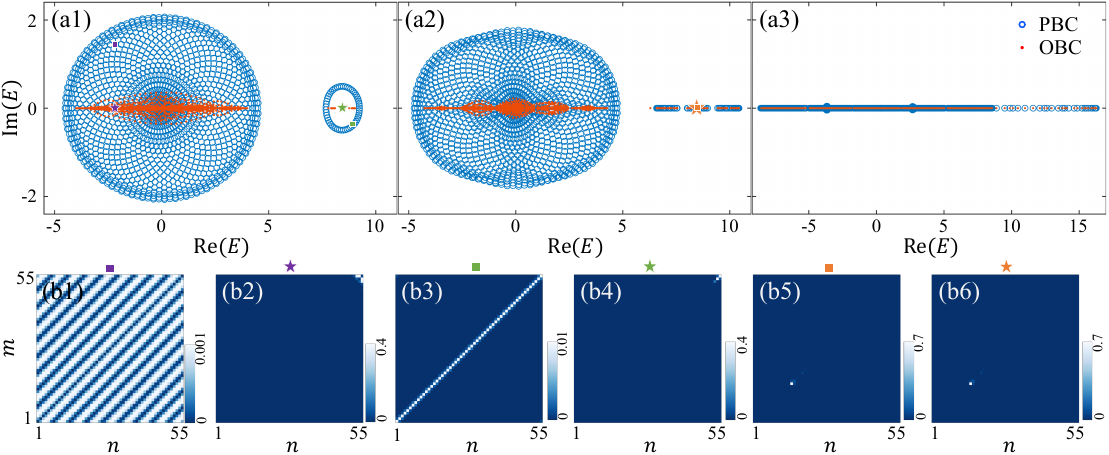}
\caption{(a1-a3) Two-particle energy spectra in the complex plane for $V=0.1$ (a1), $%
V=0.5$ (a2), and $V=2$ (a3) of the Hamiltonian (\ref{Eq1}) with PBC (blue circles)
and OBC (red dots). (b1, b2) Two-particle density distribution of the
scattering state labeled in (a1) under PBC (b1) and OBC (b2). (b3, b4)
Two-particle density distribution of the bound state states labeled in (a1)
under PBC (b3) and OBC (b4). (b5, b6) Two-particle density distribution of
the bound state states labeled in (a2) under PBC (b5) and OBC (b6). The
other parameters are given by $\alpha =0.5$, $U=8,$ and $L=55$.}
\label{fig4}
\end{figure*}

It follows that the critical value $V_{c}^{\prime }$ is indeed
lower than $V_{c}$ when $U>V_{c}$. For large $U$, $V_{c}^{\prime }\left(
U\right) $ can well capture the phase boundary of the localization
transition of the bound states (see Fig. \ref{figA1} in Appendix \ref{Ap_B}).

\subsection{Two-particle localization and skin effect}

The above results are obtained using PBC.
It is well known that the eigenstates and eigenvalues of non-Hermitian
systems with nonreciprocal hoppings exhibit extreme sensitivity to the
presence or absence of boundaries. For example, in one-dimensional systems,
most of eigenstates are accumulated on one of the edges under open boundary
conditions (OBC), while they extended into the bulk under PBC. In this
section, we investigate the effect of boundary conditions on the
two-particle energy spectra and localization properties of the nonreciprocal
interacting quasicrystal in Eq.~(\ref{Eq1}). To get a clear manifestation of the
bound states, we will focus on the case of strong interaction.

Figure~\ref{fig4} shows the two-particle energy spectra in the complex plane for
different $V$ under both PBC and OBC. To have a clear view of the complex energy spectrum and eigenstate distribution, here we choose $L=55$.
As shown in Fig.~\ref{fig4}(a1), in the extended
phase, the OBC spectrum looks drastically different from the PBC spectrum.
Note that the isolated energy loop on the right side of the energy spectrum
describes doublon states. In this case, the extended scattering and bound
states under PBC turn into skin modes under OBC, as shown in Figs.~\ref{fig4}(b1-b4).
Figure~\ref{fig4}(a2) displays the spectrum for intermediate phase. It can be seen
that, while the eigenenergies of scattering states for PBC and OBC are
separated in the complex plane, the eigenenergies of bound states for PBC
and OBC completely overlap. Moreover, these overlapping bound states are
localized whose positions are independent of the boundary condition [Figs.~\ref{fig4}(b5) and \ref{fig4}(b6)]. The results for localized phase are plotted in Fig.~\ref{fig4}(a3).
It is clear that the energy spectra under PBC and OBC are almost the same,
implying that the localized states are not sensitive to the boundary
conditions. From the above discussion, it is found that knowledge about localization properties of the eigenstates can be easily read out by comparing the corresponding complex spectrum structures under PBC and OBC. Of course, the interaction-induced mobility edge is also encoded in the spectrum structures.

\subsection{Correlated dynamics}

In the Hermitian single-particle systems, one of the key characteristic
features of the localization is that the wavepacket dynamics shows distinct
transport behaviors at different phases, which can be effectively captured
the localization property. When two particles simultaneously reside in our
nonreciprocal quasicrystal model, interactions and non-Hermiticity strongly
modify the behaviors of the system.

We study the spreading of wavepacket by preparing two bosons on the center
site of the lattice, i.e., $\left\vert \psi \left( 0\right) \right\rangle =(1/\sqrt{%
2})\hat{a}_{L/2}^{\dagger }\hat{a}_{L/2}^{\dagger }\left\vert 0\right\rangle $,
to characterize the localization of our model. Since $\hat{H}$ is
non-Hermitian here, the Schr\"{o}dinger equation itself does not preserve
the norm of $\left\vert \psi \left( t\right) \right\rangle $. To avoid an
exponential rise or decay of the wave function with time, we need to
normalize the wave function. The evolution wave function at a time interval
$dt$ is thus given by $\left\vert \psi \left( t+dt\right) \right\rangle =e^{-i%
\hat{H}t}\left\vert \psi \left( t\right) \right\rangle $ and $\left\vert
\psi \left( t+dt\right) \right\rangle =\left\vert \psi \left( t+dt\right)
\right\rangle /\left\Vert \left\vert \psi \left( t+dt\right) \right\rangle
\right\Vert $.

\begin{figure}[t]
\centering
\includegraphics[width=7.5cm]{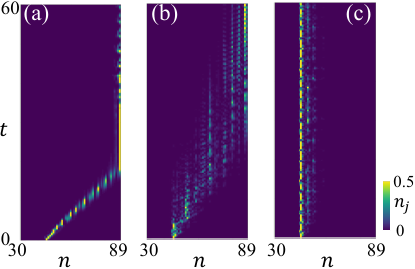}\newline
\caption{Evolution of two particles initially localized at the center site of the
lattice i.e., $\left\vert \psi \left( 0\right) \right\rangle =(1/\sqrt{%
2})\hat{a}%
_{45}^{\dagger }\hat{a}_{45}^{\dagger }\left\vert 0\right\rangle $, for $V=0.5$ (a), $V=1.5$ (b), and $V=2$ (c). The other parameters are the same as
those in Fig. \ref{fig2}.}
\label{fig5}
\end{figure}

\begin{figure*}[t]
\centering
\includegraphics[width=1.8\columnwidth]{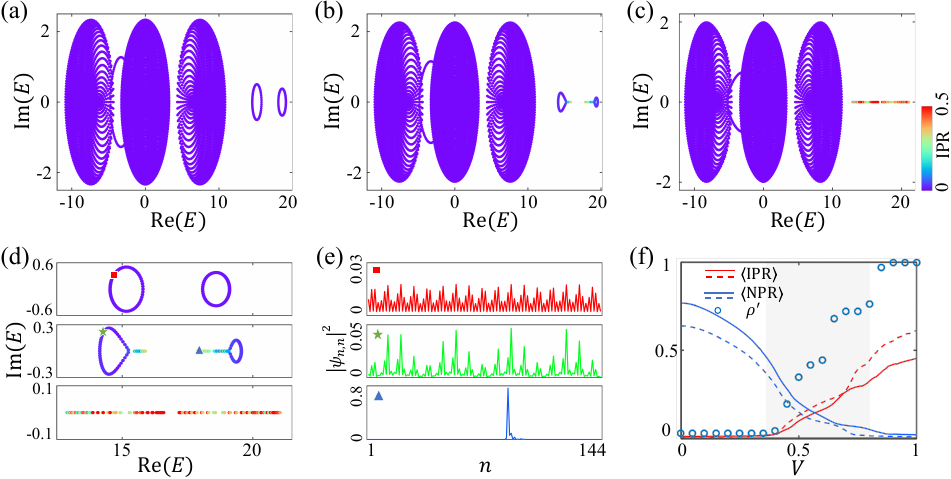}
\caption{(a-c) Two-particle energy spectra of the Hamiltonian (\ref{Eq9}) in the complex
plane for $V=0.2$ (a), $V=0.5$ (b), and $V=1$ (c). The color code indicates
the values of IPR$_{\left( l\right) }$. 
(d) The enlarged view of the bound states in (a-c). (e) Distributions $\left\vert \psi _{n,n}\right\vert ^{2}$ of the eigenstates corresponding to the eigenvalues labeled in (d). (f) The $\left\langle \text{IPR}\right\rangle $ (red
curves) and the $\left\langle \text{NPR}\right\rangle $ (blue curves) as functions of $V$, respectively. The solid and dashed lines respectively represent the numerical results obtained from Hamiltonian (\ref{Eq9}) and (\ref{Eq10}). The circles denote the results of $\rho^{\prime}$ versus $V$. The other parameters are given by $\alpha =0.5$, $\delta =4$, $U=15$, and $L=144$.}
\label{fig6}
\end{figure*}

Figure~\ref{fig5} shows the time evolution of the $n_{j}\left( t\right) =\left\langle
\psi \left( t\right) \right\vert \hat{n}_{j}\left\vert \psi \left( t\right)
\right\rangle $ for different values of $V$ under OBC. As shown in Fig.~
\ref{fig5}(a), when the two-boson wave function initialized in the extended phase, we
observe that it is ballistically transported to one side of the lattice and
then remains at the boundary, which is due to the presence of skin modes on
the corresponding boundary. In the intermediate phase [Fig.~\ref{fig5}(b)], however,
while initially localized for a short time, the two-particle wavepacket
jumps to other sites and gradually moves to the right side of the lattice.
This behavior can be understood as follows. The intermediate phase features
the coexistence of complex and real eigenenergies [Fig.~\ref{fig4}(b)] under OBC,
corresponding to skin scattering states and localized bound states,
respectively. As a result, the $\left\vert \psi (t)\right\rangle $ is
dominated by the skin modes with large imaginary part in a long-time domain. On the other hand, for the localize
phase [Fig.~\ref{fig5}(c)], the two-particle wavepacket is frozen into its initial
position without spreading.

\section{Interaction-induced mobility edge of bound states in a dimerized nonreciprocal quasicrystal}

In Sec.~\ref{SII}, we have shown that the interaction leads to a lower
threshold for the extended-localized transition of bound states compared
to scattering states, resulting in the emergence of mobility edges separating the localized bound states from the extended scattering states. In this
section, by extending the above discussions to a general quasiperiodic Hubbard
lattice with dimerized nonreciprocal hoppings, we reveal that the
bound states can form mobility edges on their own under the interplay of dimerization and interaction, displaying the coexistence of extended and localized bound states.

The Hamiltonian of the dimerized nonreciprocal interacting quasicrystal is
described as
\begin{eqnarray}
\hat{H}^{\prime } &\!=\!&\sum\nolimits_{j\in \text{odd}}^{L}\left[ J_{1}(
e^{\alpha }\hat{a}_{j+1}^{\dagger }\hat{a}_{j}+e^{-\alpha }\hat{a}%
_{j}^{\dagger }\hat{a}_{j+1} \right]   \nonumber \\
&&+\sum\nolimits_{j\in \text{even}}^{L}\left[ J_{2}( e^{\alpha }\hat{a}%
_{j+1}^{\dagger }\hat{a}_{j}+e^{-\alpha }\hat{a}_{j}^{\dagger }\hat{a}%
_{j+1}) \right] \notag \\
&&+\!\sum_{j}^{L}2V\cos \left( 2\pi \beta j\right) \hat{n}_{j}\!+\!\frac{U%
}{2}\sum_{j}^{L}\hat{n}_{j}\left( \hat{n}_{j}\!-\!1\right) .  
\label{Eq9}
\end{eqnarray}%
For convenience, we define a quantity $\delta =J_{2}/J_{1}$ to specify the
strength of hopping dimerization and set $J_{1}=1$ as the energy unit. Figures~\ref{fig6}(a-c) show the two-particle
energy spectra of Hamiltonian (\ref{Eq9}) associated with the IPR$_{(l)}$
for different $V$ with $\left\{ \alpha =0.5,\delta =4\right\} $ on the
complex plane. Here, an even Fibonacci number ($L=144$) should be taken to satisfy the PBC. We can find that, in addition to the full phases of complex and real
eigenenergies, the spectra of bound states experiencing an intermediate
region showing the coexistence of complex and real eigenenergies by
increasing the disorder strength, which are shown more clearly in Fig.~\ref{fig6}(d). Additionally, the spectrum of the bound states forms two isolated loop or two line-gapped band compared to the Hamiltonian (\ref{Eq1}) (see Fig.~\ref{fig4}).
Figure~\ref{fig6}(e) displays the typical density distributions $\psi _{n,n}$ of the
states labeled in Fig.~\ref{fig6}(d). It can be seen that the bound states with
complex (real) eigenenergies are extended (localized).

To illustrate the localization transitions of bound states clearer, we define
the IPR and NPR of the bound states in the doublon subspace, i.e., IPR$%
_{\left( l\right) }=\sum\nolimits_{n=1}^{L}\left\vert \psi _{n,n}^{\left(
l\right) }\right\vert ^{4}$ and NPR$_{\left( l\right) }=$IPR$_{\left(
l\right) }^{-1}/L$ with $l=\{1,\cdots ,L\}$. We further average the IPR$_{(l)}$ and NPR$_{(l)}$ over
all bound states to obtain $\left\langle \text{IPR}\right\rangle $ and $%
\left\langle \text{NPR}\right\rangle $, where the symbol $\left\langle
\cdots \right\rangle $ indicates the averaged value. Figure~\ref{fig6}(f) plots the $%
\left\langle \text{IPR}\right\rangle $ (red-solid line) and the $\left\langle \text{NPR}%
\right\rangle $ (blue-solid line) as functions of $V$ for Hamiltonian (\ref{Eq9}). An intermediate phase (shaded region) where both $\left\langle \text{IPR}\right\rangle $ and $%
\left\langle \text{NPR}\right\rangle $ are finite is clearly identified, implying the coexistence of
extended and localized bound states. That is to say, a mobility edge of bound states is formed in this parameter region. Similarly, the localization transition of bound states coincides with a complex-real transition in the spectrum. To this end, we analyse the behaviour of the fraction of real eigenvalues $\rho' =L_{0}/L$, where $L_{0}$ counts the number of bound states having real eigenenergies in the spectrum. As presented in Fig.~\ref{fig6}(f), there also exists a region where the value of $\rho'$ is in between zero and one, which indicates the presence of both real and imaginary eigenenergies in the spectrum. Its boundaries are consistent with the intermediate phase with mobility edges.

Using the perturbation theory presented in Appendix \ref{Ap_A}, the effective doublon Hamiltonian of Hamiltonian (\ref{Eq9}) in the large on-site interaction $U$ takes the form%
\begin{eqnarray}
\hat{H}_{d}^{\prime } &=&\sum\nolimits_{j\in \text{odd}}^{L}\left[ \frac{%
2J_{1}^{2}}{U}\left( e^{2\alpha }\hat{d}_{j+1}^{\dagger }\hat{d}%
_{j}+e^{-2\alpha }\hat{d}_{j}^{\dagger }\hat{d}_{j+1}\right) \right]  
\nonumber \\
&&+\sum\nolimits_{j\in \text{even}}^{L}\left[ \frac{2J_{2}^{2}}{U}\left(
e^{2\alpha }\hat{d}_{j+1}^{\dagger }\hat{d}_{j}+e^{-2\alpha }\hat{d}%
_{j}^{\dagger }\hat{d}_{j+1}\right) \right]   \nonumber \\
&&+\sum\nolimits_{j}^{L}4V\cos \left( 2\pi \beta j\right) \hat{n}_{j}^{d}.
\label{Eq10}
\end{eqnarray}
Equation (\ref{Eq10}) describes a dimerized nonreciprocal lattice of doublons with quasiperiodic disorder. Notably, it has been demonstrated that the dimerization ($\delta\neq1$) can lead to the single particle
 mobility edge \cite{PRL106803,NJP123048,PRB1054204,PRB054307}. In Fig.~\ref{fig6}(f), we further plot $\left\langle \text{IPR}\right\rangle $ (red-dashed line) and $\left\langle \text{NPR}\right\rangle $ (blue-dashed line) as a function of $V$ based on Hamiltonian $\hat{H}_{d}^{\prime }$. One can see that the boundary of the intermediate phase is in good agreement with those of $\hat{H}^{\prime }$.

\section{POSSIBLE EXPERIMENTAL IMPLEMENTATION}

\begin{figure}[t]
\centering
\includegraphics[width=8cm]{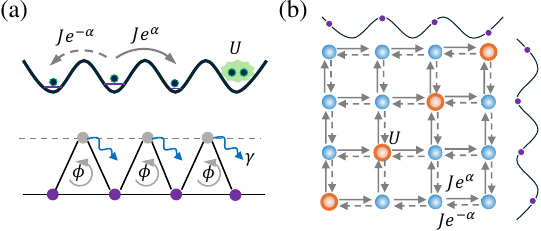}\newline
\caption{(a) Top: sketch of the nonreciprocal Hubbard model with a quasiperiodic on-site potential. Bottom: sketch of a dissipative Aharonov-Bohm chain with a synthetic magnetic flux $\phi$ through each ring. The gray dots indicate dissipative sites with rate $\gamma$. (b) Sketch of the mapping onto a two-dimensional nonreciprocal quasicrystal.}
\label{EXPP}
\end{figure}

The key points of the nonreciprocal interacting quasicrystal are to realize the asymmetric hoppings and interactions, as shown in the top panel of Fig.~\ref{EXPP}(a). Ultracold atoms provide a promising platform to achieve these. Note that, instead of directly realizing asymmetric hopping, a nonreciprocal lattice with non-Hermitian skin effect can be achieved by the interplay of dissipation and synthetic magnetic flux \cite{PRL186802,PRA1302,PRA0202,PRA3329}, as shown in the bottom panel of Fig.~\ref{EXPP}(b). Along this line, the non-Hermitian skin effect has been demonstrated in ultracold quantum gases \cite{PRL0401}. The quasiperiodic potential can be implemented through the superposition of a primary lattice and a weaker incommensurate lattice, i.e., bichromatic optical lattice \cite{PRL1604}. In addition, the incommensurate modulation can be also realized in momentum-state lattice of quantum gas, where the on-site potential of each site can be controlled by tuning the frequency of a corresponding frequency component of the multifrequency beam \cite{PRX045,PRL3401}.

As demonstrated in Eq.~(\ref{Eq3}), the two-particle nonreciprocal quasicrystal in one dimension can be mapped to a single particle in two-dimensional nonreciprocal quasicrystal, as sketched in Fig.~\ref{EXPP}(b). Such mapping provides another convenient way to realize our results in various platforms. Strikingly, because of the wide choice of circuit components, electric circuit is an excellent platform for engineering non-Hermitian Hamiltonians. For the lattice in Fig.~\ref{EXPP}(b), the nonreciprocal hoppings can be implemented through an impedance converter with current inversion (INIC) \cite{PRB195131,NC98}, and the on-site quasiperiodic potential can be realized by grounding the nodes with position-dependent capacitors. In the experiment, we can acquire the eigenvalues and eigenstates of the system under PBC and OBC by measuring the voltage response at each node to a local current input. By tuning the capacitance of each node, the system can switch between different localization phases.

\section{Conclusions}

In summary, we have investigated localization features of two interacting bosons in one-dimensional nonreciprocal quasicrystals. Considering a nonreciprocal quasiperiodic AAH model with on-site two-body interaction, we found that the interaction can lead to an intermediate phase with mobility edges. By analyzing the spatial distribution of eigenstates and the corresponding eigenenergies, we showed that such a mobility edge is caused by the fact that the bound states display a much lower threshold for spectral and delocalization-localization transitions than other scattering states. We have also studied the impacts of boundary conditions on the two-particle localization properties. It is found that the two-particle extended states turn into skin modes under OBC, whereas the two-particle localized states are robust to different boundary conditions. The correlated dynamics was also demonstrated to characterize the localization transitions. Finally, we revealed that the bound states can form a mobility edge on their own by introducing a dimerized nonreciprocal quasicrystal. Our results are an important step towards understanding the interplay between quasiperiodicity and interactions in non-Hermitian systems.

\section{Acknowledgements}

This work is supported by the National Key R\&D Program of China (Grant
 No.~2022YFA1404500, No.~2021YFA1400900), Cross-disciplinary Innovative Research Group
 Project of Henan Province (Grant No. 232300421004), the National Natural Science Foundation of China (Grant No.~12125406, No.~12374312, No.~12474340, and No.~12404372), and China Postdoctoral Science Foundation (Grant No. GZB20240684).

\appendix

\begin{figure}[t]
\centering
\includegraphics[width=6.5cm]{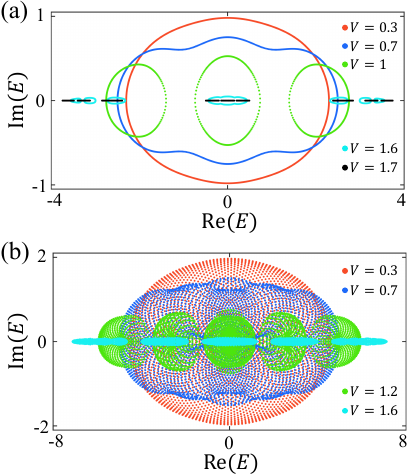}\newline
\caption{(a) Single-particle energy spectra of Hamiltonian~(\ref{Eq1}) for different $V$ in the complex plane with $L=610$. (b) Two-particle complex energy spectra of Hamiltonian~(\ref{Eq1}) for different $V$ in the noninteracting limit. The other parameters are the same as those in Fig.~\ref{fig1}.}
\label{STPT}
\end{figure}

\section{Noninteracting two-particle energy spectrum for different $V$}
\label{Ap}

In the noninteracting limit, the two-particle eigenenergies are rooted in the single-particle spectral properties. As shown in Fig.~\ref{STPT}(a), we present the single-particle energy spectra of Hamiltonian~(\ref{Eq1}) for different $V$ in the complex plane. It can be seen that the spectrum forms a single loop structure for a small value of $V$. With the increase of $V$, the spectrum splits into multiple separated loops. The number of split loops increases as the critical point $V_c$ is approached, and the radius of the loops shrinks as the energy spectrum becomes fully real. In this context, the noninteracting two-particle energy spectrum also displays different structures at different $V$. Figure~\ref{STPT}(b) displays the two-particle energy spectra with various V. Clearly, the two-particle energy spectrum forms one filled loop when $V$ is small. As $V$ is further increased yet remains below the critical value $V_c$, the spectrum exhibits multiple clusters, corresponding to the case of multiple loops of the single-particle energy spectrum.

\section{Finite size analysis}
\label{AFS}

\begin{figure}[t]
\centering
\includegraphics[width=8cm]{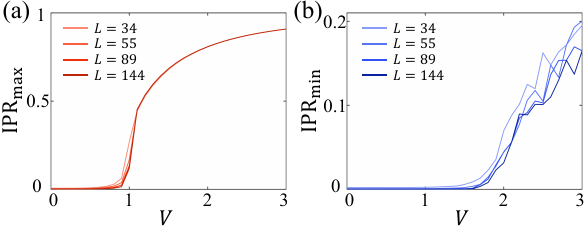}\newline
\caption{The IPR$_{\max }$ (a) and IPR$_{\min}$ (b) as functions of $V$ with $L=34$, $55$, $89$, and $144$. The other parameters are the same as those in Fig. \ref{fig2}.}
\label{FINT}
\end{figure}

To confirm that the two-particle localization transitions are independent of the system size, we numerically calculate the IPR$_{\max }$ and IPR$_{\min}$ as functions of $V$ for different $L$, as shown in Fig.~\ref{FINT}. It is clearly shown that the sharpness of the change of IPR$_{\max }$ and IPR$_{\min}$ become larger with the increase of the system size, which indicates the stability of the mobility edge region.

\section{Derivation of the effective doublon Hamiltonian}
\label{Ap_A}

In this appendix, we provide details on the derivation of the effective
Hamiltonian for the two-particle bound state in the strong-interaction limit
$U\gg J$, $V$. For the purpose of the derivation of the effective
Hamiltonian in the doublon subspace, it is useful to write the Hamiltonian $%
\hat{H}=\hat{H}_{0}+\hat{H}_{\text{int}}$ as the sum of the noninteracting
part $\hat{H}_{0}$ and interacting term $\hat{H}_{\text{int}}=\frac{U}{2%
}\sum\nolimits_{j}^{L}\hat{n}_{j}\left( \hat{n}_{j}-1\right) $. We are going
to treat $\hat{H}_{0}$ as a perturbation.

The interaction Hamiltonian $\hat{H}_{\text{int}}$ has two distinct types of
degenerate eigenstates --- the dounblon states $\left\vert d_{j}\right\rangle =(1/\sqrt{%
2})\hat{a}_{j}^{\dagger }\hat{a}_{j}^{\dagger }\left\vert 0\right\rangle =%
\hat{d}_{j}^{\dagger }\left\vert 0\right\rangle $ with energies $E_{d}=U$, and
the other single-occupied states $\left\vert s_{j,j^{\prime }}\right\rangle =%
\hat{a}_{j}^{\dagger }\hat{a}_{j^{\prime }}^{\dagger }\left\vert
0\right\rangle $ with energies $E_{s}=0$.
Based on the second-order degenerate perturbation theory, the nonzero matrix
elements of the effective doublon Hamiltonian $\hat{H}_{d}$ are given by%
\begin{eqnarray}
\left\langle d\right\vert \hat{H}_{d}\left\vert d^{\prime }\right\rangle 
&=&E_{d}\delta _{dd^{\prime }}+\left\langle d\right\vert \hat{H}%
_{0}\left\vert d^{\prime }\right\rangle   \nonumber \\
&&+\frac{1}{2}\sum\nolimits_{s}\left\langle d\right\vert \hat{H}%
_{0}\left\vert s\right\rangle \left\langle s\right\vert \hat{H}%
_{0}\left\vert d^{\prime }\right\rangle \notag \\
&&\times \left( \frac{1}{E_{d}-E_{s}}+\frac{1}{E_{d^{\prime }}-E_{s}}\right). 
\label{EqA1}
\end{eqnarray}%
\newline
The matrix elements in Eq.~(\ref{EqA1}) give the effective doublon hopping if $d\neq
d^{\prime }$ and effective on-site potential for $d=d^{\prime }$. 
Substituting the $\hat{H}_{0}$ in Hamiltonian (\ref{Eq1}) into Eq.~(\ref{EqA1}), we can
obtain the corresponding effective on-site potentials and nonreciprocal hoppings, which are respectively given by
\begin{eqnarray}
\left\langle d_{j}\right\vert \hat{H}_{d}\left\vert d_{j}\right\rangle
&=&\left\langle d_{j}\right\vert 2V\cos (2\pi \beta j)\hat{n}_{j}\left\vert
d_{j}\right\rangle +U  \notag \\
&&+\frac{2J^{2}}{U}\sum_{s,i,i^{\prime }}\left\langle d_{j}\right\vert \hat{a%
}_{i+1}^{\dagger }\hat{a}_{i}\left\vert s\right\rangle \left\langle
s\right\vert \hat{a}_{i^{\prime }}^{\dagger }\hat{a}_{i^{\prime
}+1}\left\vert d_{j}\right\rangle   \notag \\
&=&4V\cos (2\pi \beta j)+U+\frac{4J^{2}}{U},
\label{A2}
\end{eqnarray}

\begin{figure}[t]
\centering
\includegraphics[width=6.5cm]{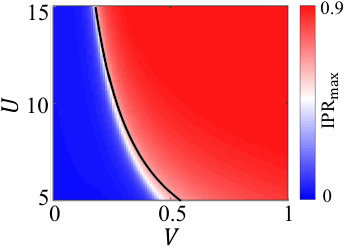}\newline
\caption{Phase diagrams of IPR$_{\max }$ in the $V$-$U$ plane for Hamiltonian (\ref{Eq1}). The black curve is obtained from Eq. (\ref{Eq8}). The other parameters are the same as those in Fig. \ref{fig2}.}
\label{figA1}
\end{figure}

\begin{eqnarray}
\left\langle d_{j\!+\!1}\right\vert \!\hat{H}_{d} \!\left\vert d_{j}\right\rangle &\!\!=\!\!&
\frac{1}{U}\sum\nolimits_{s}\left\langle d_{j+1}\right\vert \hat{H}%
_{0}\left\vert s\right\rangle \left\langle s\right\vert \hat{H}%
_{0}\left\vert d_{j}\right\rangle   \notag \\
&\!=\!&\frac{J^{2}e^{2\alpha }}{U}\sum_{s,i,i^{\prime }}\left\langle
d_{j\!+\!1}\right\vert \! \hat{a}_{i\!+\textbf{}1}^{\dagger }\hat{a}_{i} \!\left\vert
s\right\rangle \! \left\langle s\right\vert \! \hat{a}_{i^{\prime }\!+\!1}^{\dagger }%
\hat{a}_{i^{\prime }} \!\left\vert d_{j}\right\rangle   \notag \\
&=&\frac{2J^{2}e^{2\alpha }}{U},
\end{eqnarray}

and
\begin{eqnarray}
\left\langle d_{j}\right\vert \!\hat{H}_{d}\!\left\vert d_{j\!+\!1}\right\rangle  &\!=\!&%
\frac{J^{2}e^{2\alpha }}{U}\!\sum_{s,i,i^{\prime }}\left\langle
d_{j}\right\vert\! \hat{a}_{i}^{\dagger }\hat{a}_{i\!+\!1} \!\left\vert
s\right\rangle \!\left\langle s\right\vert \! \hat{a}_{i^{\prime }}^{\dagger }%
\hat{a}_{i^{\prime }+1} \!\left\vert d_{j\!+\!1}\right\rangle   \notag \\
&=&\frac{2J^{2}e^{-2\alpha }}{U}.
\label{A4}
\end{eqnarray}
According to Eqs.~(\ref{A2})-(\ref{A4}), we can obtained the effective doublon Hamiltonian (\ref{Eq7}) by neglecting a dispensable constant. The effective doublon Hamiltonian of the dimerized nonreciprocal interacting quasicrystal can be achieved in the same way.

\section{Localization transition of bound states in the strong interaction}
\label{Ap_B}

Figure~\ref{figA1} presents the phase diagram of IPR$_{\max }$ of Hamiltonian (\ref{Eq1}),
where the black curve is given by Eq. (\ref{Eq8}). This figure indicates that the
localization transition of the bound states can be well described by the
effective doublon Hamiltonian (\ref{Eq7}) in the strong-interaction limit $U\gg J, V$.


\end{document}